# A Jacobian Separable 2-D Finite-Element Method for Electromagnetic Waveguide Problems

Ehsan Khodapanah

*Abstract*−**We propose an efficient finite-element analysis of the vector wave equation in a class of relatively general curved polygons. The proposed method is suitable for an accurate and efficient calculation of the propagation constants of waveguides filled with pieces of homogeneous materials. To apply the method, we first decompose the 2-D problem domain into a set of curved polygons of a specific characteristic. Then we divide every polygon into a set of triangular elements with two straight edges. Finally, we introduce a set of hierarchical mixed-order curl-conforming vector basis functions inside every triangular element to discretize the vector wave equation. The advantages of the method are as follows. The curved boundaries of the elements are modeled exactly and hence there is no approximation in the geometrical modeling. 2-D integrals of the matrix elements are reduced to 1-D integrals. Therefore, the matrix filling can be performed very fast. Total number of elements due to the discretization of a given domain is very small, and hence the discretization of the problem domain takes up a very small percentage of the total computational time. We validate the method by comparing the results with those of Ansoft HFSS simulator and investigate the accuracy and efficiency of the method through some numerical examples.**

*Index terms*−**Finite-element method (FEM), inhomogeneous waveguides, curved polygonal elements, hierarchical basis functions.**

## I. INTRODUCTION

Finite-element method (FEM) is a powerful and accurate numerical algorithm to solve scalar and vector differential equations in physics and engineering. Its formulation and application to a variety of electromagnetic field problems can be found in many books (e.g., [1]-[3]). The method has been successfully applied to analyze 2-D waveguide eigenvalue problems as a special class of vector electromagnetic field problems [4]-[9].

Higher order FEMs are more popular due to their higher accuracy for a given number of unknowns. However, the word "order" in this context has a two-fold meaning. The order may refer to the geometrical order of the cells that are used to decompose the computational domain or to the order of the expansion that is used to represent the unknown field inside the cells. Let us consider the second case. Most of the higher order bases that are used in the computational electromagnetics are constructed based on the criteria that have been introduced by Nedelec [10], [11]. For example, higher order interpolatory bases, which span the incomplete-order (reduced-gradient) polynomial spaces of Nedelec, are constructed in [12] and higher order hierarchical bases for triangular and tetrahedral cells, which span the complete-order polynomial spaces of Nedelec, are introduced in [13]. It should be mentioned that the hierarchical bases are more popular due to their ability to perform p-adaptions. Higher order hierarchical polynomial bases for curved quadrilateral and hexahedral elements are introduced in [14] and applied to the analysis of waveguide problems [15] and Maxwell's equations directly in the time domain [16]. Higher order hierarchical bases utilizing orthogonal polynomials to improve the condition number of the finite-element matrices are constructed in [17] for triangular and tetrahedral cells and in [18] for quadrilateral and brick cells. In [19], hierarchical bases, which span both incomplete- and complete-order spaces of Nedelec on tetrahedral cells, are constructed in a way that the higher order bases are orthogonal to the lower order ones and hence the basis are suitable for error estimations in multilevel solvers.

When a finite-element cell is defined by a set of independent variables, each of which varies in a constant range (i.e., there is a mapping from a rectangular or brick element to the original finite element cell), it is possible to construct higher order interpolatory or hierarchical mixed-order curl-conforming vector basis functions that span the null-space of the curl operator correctly and hence avoid the occurrence of non-zero spurious solutions. These bases, which do not necessarily span the polynomial spaces of Nedelec, are used in [14], [20], [21] for constructing higher order vector basis functions according to the tensor product of the scalar 1-D functions of the variables defining the cell. These bases have the advantages that they provide anisotropic expansions in the different directions inside the elements and also have simple explicit expressions for arbitrary orders. When these bases are isotropically constructed in the rectangular or brick elements,

The author is with the Department of Electrical and Computer Engineering, University of Tabriz, Tabriz, Iran (e-mail: ekhodapanah@ tabrizu.ac.ir).

they span exactly the mixed-order Nedelec spaces, however, they can also be constructed on triangular elements if we adopt the parametric definitions given in [22] (i.e., the Duffy transformation).

Now, let us consider the geometrical order. The finite-element bases are first constructed inside a standard straight element (a reference element) and then a specific mapping is used to transform the basis functions into the curved physical element (e.g., see [1], [14]). An important note, which should be considered, regarding to this transformation, is that the Jacobian of this transformation (the Jacobian of the covariant matrix) appears in the denominator of the integrands of the stiffness and mass matrices and couples the integration variables together. Therefore, fully 2-D integrations (in the 2-D problems) should be carried out for the evaluation of the elemental matrices. It has been found that when a polar angle is used to define the geometry, the Jacobian of the transformation will be separated in the special ring type elements with one geometrical degree of freedom (i.e., when the outer and inner boundaries are linearly dependent) [23]. For a general ring type element with two degrees of freedom (i.e., when the outer and inner boundaries are independent) the Jacobian is not separated but appears as a sum of two separable terms. Therefore, multiplying the weighting functions by appropriate powers of the Jacobian term converts this term from the denominator to the numerator in the integrands, and hence 2-D integrals of the matrix elements are reduced to 1-D ones [24].

In this paper, we present a specific decomposition of a 2-D domain to model the curved boundaries exactly. Then we make use of a polar angle to define a set of tensor-product curl-conforming vector bases inside the curved elements. The method leads to 1-D integrals for elemental matrices and hence is very efficient in the matrix filling procedure. The proposed method then is applied to analyze a regular waveguide and two singular waveguides and the accuracy and efficiency of the method are investigated.

The rest of the paper is organized as follows. In section II, the finite-element formulation of a 2-D waveguide problem based on the new domain decomposition and basis functions is introduced. A method for the determination of the common vertices of the curved polygons is given in section III. In section IV, the method of section II is applied to analyze a regular waveguide and two singular waveguides and the efficiency and accuracy of the method are investigated. Finally in section V, a conclusion is given.

## II. Theory

In order to describe our efficient finite-element discretization, we consider a uniform waveguide in the longitudinal $z$ direction, including an inhomogeneous material with a piecewise constant inhomogeneity in the transverse $x - y$ plane and bounded by a perfect magnetic conducting (PMC) wall. To implement the proposed method, we first divide the 2-D problem domain into a number of curved polygons of a specific characteristic; there exists a point inside every polygon such that the closed boundary of the polygon can be represented by a piecewise smooth function of polar angle in the local polar coordinate system centered at this point. Then each polygon is subdivided into a number of triangular elements with a common vertex located at the origin of the local coordinate system inside the polygon. Each triangular element has two straight edges that are located inside the polygon and one curved edge that is shared by the boundary of the polygon and represented by a smooth function in the local polar coordinate system of the same polygon. The result of this subdivision is the decomposition of the domain into a set of triangular elements. Two adjacent triangular elements may be connected through a straight line inside a polygon or a curved line at the common boundary of two adjacent polygons. We consider the latter case and illustrate a typical example of this case in Fig. 1. In Fig. 1, the points $(x_p, y_p)$ and $(x_a, y_a)$ represent the origins of the local coordinate systems corresponding to the two triangular elements and $(\rho, \varphi)$ and $(\rho', \varphi')$ are the polar coordinates of a point in the $p$ and $a$ systems, respectively. Also in Fig. 1, $\rho = \rho_1(\varphi)$ and $\rho' = \rho_2(\varphi')$ are the definitions of the common boundary in the $p$ and $a$ coordinate systems, respectively, where $\rho_1(\varphi)$ and $\rho_2(\varphi')$ are smooth functions of $\varphi$ and $\varphi'$, respectively, according to our assumption about the unique feature of the polygonal elements. We assume that the mathematical expression of the common boundary of the two elements is given explicitly in the local coordinate system of one of the two elements. This element is called a principal element while the other one is called an associate element. The subscripts $p$ and $a$ in Fig. 1, denote the principal and associate, respectively.

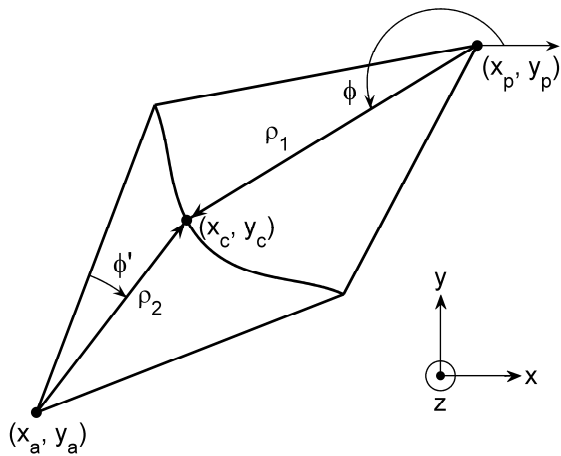

Fig. 1. Geometry of the principle and associate elements.

The principal element can be defined by two variables $u$ and $\varphi$ as $\rho = u\rho_1(\varphi)$ for $\varphi_1 \le \varphi \le \varphi_2$ and $0 \le u \le 1$. According to this parametric definition, mixed-order vector basis functions in the principal element of Fig. 1 can be easily constructed based on a tensor product of scalar 1-D basis functions of the $u$ and $\varphi$ variables. In other words, the electric field vector in the principal element is approximated by

$$\vec{e}_t = \sum_{m=0}^{M_u-1} \sum_{n=0}^{M_\varphi} a_{mn} U_m(2u-1) T_n\left(1 - 2\frac{\varphi - \varphi_2}{\varphi_1 - \varphi_2}\right) \nabla u + \sum_{m=1}^{M_u} \sum_{n=0}^{M_\varphi-1} b_{mn} T_m(2u-1) U_n\left(1 - 2\frac{\varphi - \varphi_2}{\varphi_1 - \varphi_2}\right) \frac{\hat{\varphi}}{\rho}$$

$$e_z = \sum_{m=0}^{M_u} \sum_{n=0}^{M_\varphi} c_{mn} T_m(2u-1) T_n\left(1 - 2\frac{\varphi - \varphi_2}{\varphi_1 - \varphi_2}\right) \qquad (1)$$

where $T$ represents the rearranged Chebyshev polynomials of the first kind as indicated in [24] ,$U$ is the Chebyshev polynomial of the second kind, and $a_{mn}$, $b_{mn}$, and $c_{mn}$ are the unknown field coefficients to be determined. Substituting (1) into the vector wave equation [24] and applying the Galerkin testing procedure, we obtain the following explicit expressions for the matrix elements in the principal element

$$A_{uu}^{m_1 m_2 n_1 n_2} = \frac{1}{\mu_r} U_{m_1 m_2}^1 \Phi_{n_1 n_2}^1 + \left(\frac{\beta^2}{\mu_r} - k_0^2 \varepsilon_r\right) U_{m_1 m_2}^2 \Phi_{n_1 n_2}^2 \qquad (2)$$

$$A_{u\varphi}^{m_1 m_2 n_1 n_2} = -\frac{1}{\mu_r} U_{m_1 m_2}^3 \Phi_{n_1 n_2}^3 + \left(\frac{\beta^2}{\mu_r} - k_0^2 \varepsilon_r\right) U_{m_1 m_2}^4 \Phi_{n_1 n_2}^4 \qquad (3)$$

$$A_{uz}^{m_1 m_2 n_1 n_2} = \frac{-j\beta}{\mu_r}\left(U_{m_1 m_2}^5 \Phi_{n_1 n_2}^2 + U_{m_1 m_2}^4 \Phi_{n_1 n_2}^5\right) \qquad (4)$$

$$A_{\varphi\varphi}^{m_1 m_2 n_1 n_2} = \frac{1}{\mu_r} U_{m_1 m_2}^6 \Phi_{n_1 n_2}^6 + \left(\frac{\beta^2}{\mu_r} - k_0^2 \varepsilon_r\right) U_{m_1 m_2}^7 \Phi_{n_1 n_2}^7 \qquad (5)$$

$$A_{\varphi z}^{m_1 m_2 n_1 n_2} = \frac{-j\beta}{\mu_r}\left(U_{m_1 m_2}^8 \Phi_{n_2 n_1}^4 + U_{m_1 m_2}^7 \Phi_{n_1 n_2}^8\right) \qquad (6)$$

$$A_{zz}^{m_1 m_2 n_1 n_2} = -\frac{1}{\mu_r}\left(U_{m_1 m_2}^9 \Phi_{n_1 n_2}^2 + U_{m_1 m_2}^7 \Phi_{n_1 n_2}^9 + U_{m_1 m_2}^8 \Phi_{n_2 n_1}^5 + U_{m_2 m_1}^8 \Phi_{n_1 n_2}^5\right) + k_0^2 \varepsilon_r U_{m_1 m_2}^{10} \Phi_{n_1 n_2}^{10} \qquad (7)$$

$$A_{zu}^{m_1 m_2 n_1 n_2} = A_{uz}^{m_2 m_1 n_2 n_1} \qquad (8)$$

$$A_{\varphi u}^{m_1 m_2 n_1 n_2} = A_{u\varphi}^{m_2 m_1 n_2 n_1} \qquad (9)$$

$$A_{z\varphi}^{m_1 m_2 n_1 n_2} = A_{\varphi z}^{m_2 m_1 n_2 n_1} \qquad (10)$$

where the U and $\Phi$ sub-matrices are defined in the Appendix. It is not difficult to show that the construction of the curl-conforming vector basis functions can be made easily if the parametric definitions of the elements are performed in a way that the common boundary of two adjacent elements is represented by a same parameter from both elements sharing that boundary [14]. This fact is considered as a base for the construction of the curl-conforming vector basis functions in the associate element. The fact that the $\rho_1(\varphi)$ and $\rho_2(\varphi')$ are smooth functions of $\varphi$ and $\varphi'$, respectively, implies that there is a one to one correspondence between $\varphi$ and $\varphi'$ and hence we can write $\varphi' = f(\varphi)$, where f is a smooth function of $\varphi$. Therefore, instead of using the angular variable of the associate element, $\varphi'$, we make use of the angular variable of the principal element, $\varphi$, to define the associate element as $u = \rho'/\rho_2(\varphi')$ and $\varphi' = f(\varphi)$ for $\varphi_1 \leq \varphi \leq \varphi_2$ and $0 \leq u \leq 1$ and approximate the unknown vector field in terms of the following curl-conforming vector basis functions of $u$ and $\varphi$ in the associate element

$$\vec{e}_t = \sum_{m=0}^{M_u-1} \sum_{n=0}^{M_\varphi} a'_{mn} U_m(2u-1) T_n\left(1 - 2\frac{\varphi - \varphi_2}{\varphi_1 - \varphi_2}\right) \nabla u + \sum_{m=1}^{M_u} \sum_{n=0}^{M_\varphi-1} b'_{mn} T_m(2u-1) U_n\left(1 - 2\frac{\varphi - \varphi_2}{\varphi_1 - \varphi_2}\right) \frac{1}{f'(\varphi)} \frac{\widehat{\varphi'}}{\rho'}$$

$$e_z = \sum_{m=0}^{M_u} \sum_{n=0}^{M_\varphi} c'_{mn} T_m(2u-1) T_n\left(1 - 2\frac{\varphi - \varphi_2}{\varphi_1 - \varphi_2}\right) \qquad (11)$$

where $\widehat{\varphi'}$ is the unit vector in the direction of the $\varphi'$. The matrix elements for the associate element are the same as those given in (2)-(10) except that the $\Phi$ sub-matrices must be redefined as those given in the Appendix.

It is clear from (2)-(10) that the matrix elements are reduced to 1-D integrals with a very small variety confirming that this reduction of the integral dimension is efficient. This phenomenon which has been called a numerical separation in [24] plays an important role in reducing the matrix filling time in the proposed finite-element method.

### III. Determination of the Common Vertices of the Curved Polygons

A point inside a curved polygon is a common vertex if every semi-infinite line starting from this point intercepts the boundary of the polygon at only one point with a non-zero angle. We assume that the boundary of the polygon is composed of a number of curved edges, each of which is represented by $g(x, y) = 0$ in the global Cartesian coordinate system where $g(x, y)$ has continuous first derivatives with respect to the $x$ and $y$ on the edge and hence there exists a unique tangent at each point of the edge. In order to determine the common vertex of the polygon, we first consider each curved edge separately and determine the loci of the common vertices for this edge. The loci of the common vertices of the polygon are then determined as the common area among the locis of all the edges.

For a convex or concave edge, the loci of the common vertices is determined by drawing the two tangent lines at the two endpoints of the edge and picking up the shaded area as shown in Fig. 2a. For an oscillatory edge, it is always possible to divide the edge into pieces of convex and concave curves, which are connected to each other smoothly. By applying the foregoing procedure to each piece and selecting the common loci of all the pieces, one can determine the loci of common vertices for the oscillatory edge as shown in Fig. 2b. As mentioned previously, the loci of the common vertices includes a set of points in the polygon with the characteristic that the boundary of the polygon can be represented by a piecewise smooth function of the polar angle in the local polar coordinate system centered at each of these points. If we denote the coordinates of a common vertex by $(x_a, y_a)$ the mathematical expression for a curved edge of the polygon in the local polar coordinate system is given by $g(x_a + \rho\cos(\varphi), y_a + \rho\sin(\varphi)) = 0$ or equivalently by $\rho = \rho_1(\varphi)$. However, explicit expression for $\rho_1(\varphi)$ is not known in general. In this paper, we assume that the mathematical expression for every curved edge in the problem domain is known explicitly in a local polar coordinate system centered at a specific point. If this point, which is denoted by $(x_p, y_p)$, is a common vertex, its corresponding edge is the curved edge of a principal element. Otherwise, an appropriate common vertex should be obtained using the method described above. In this later case, the edge is the curved edge of an associate element, which is associated to the point $(x_p, y_p)$ [see Fig. 2c].

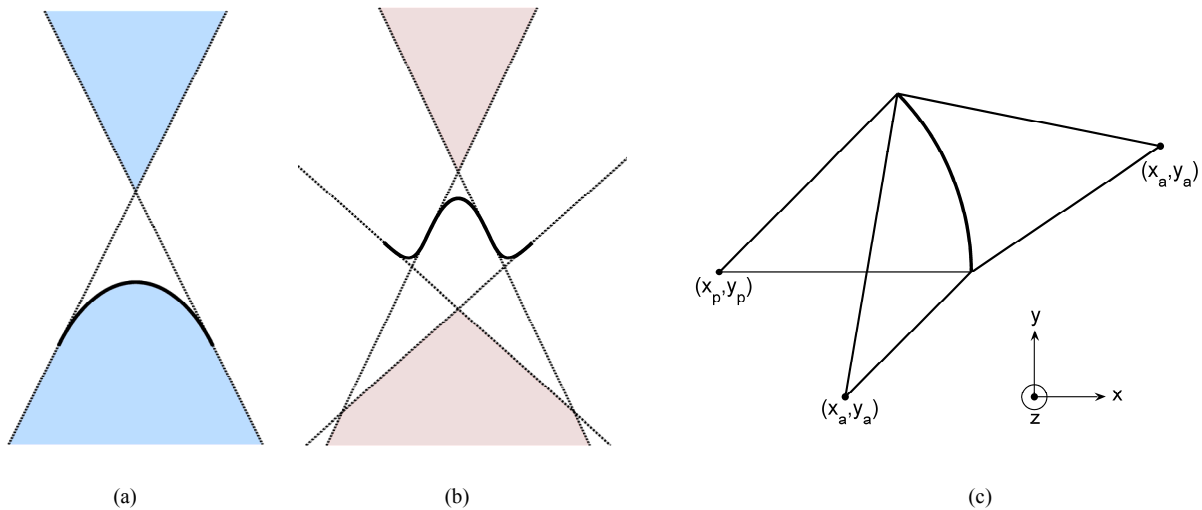

(a)                    (b)                    (c)

Fig. 2. (a) Geometry of a convex or concave edge (b) Geometry of an oscillatory edge (c) The curved edge is defined explicitly in the polar coordinates centered at $(x_p, y_p)$. If $(x_p, y_p)$ is a common vertex, the edge is the curved edge of a principal element. If $(x_a, y_a)$ is a common vertex, the edge is the curved edge of an associate element, which is associated to the point $(x_p, y_p)$.

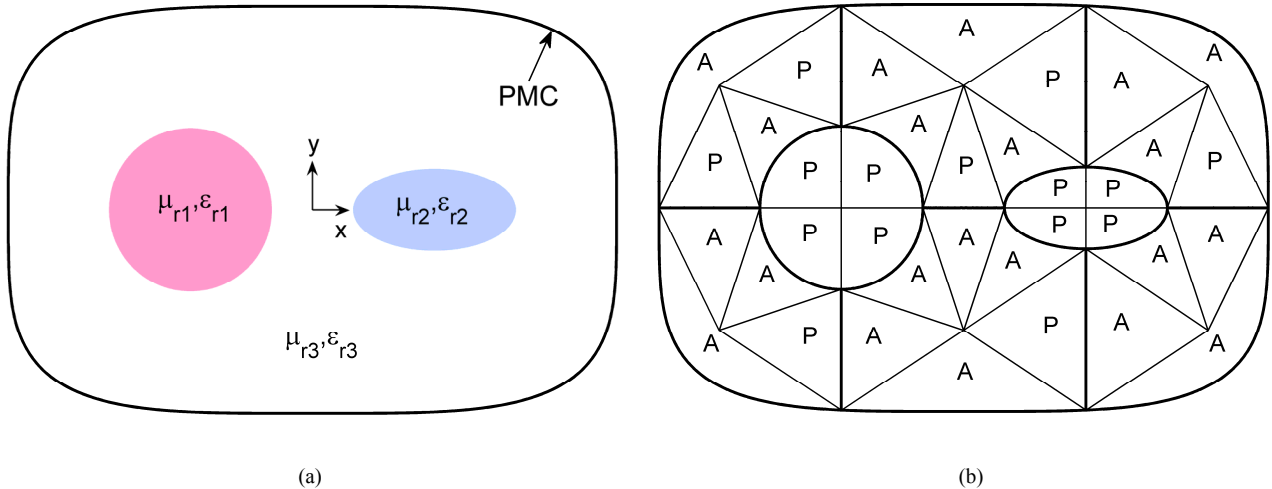

(a)

(b)

Fig. 3. (a) Geometry of the regular waveguide that is considered in section IV. The circle, ellipse, and outer boundary are defined by $(x + 0.6)^2 + y^2 = 0.4^2$, $(x - 0.6)^2 + (2y)^2 = 0.4^2$, and $0.2x^4 + y^4 = 1$, respectively, and $\varepsilon_{r1} = \varepsilon_{r2} = 4$, $\varepsilon_{r3} = 1$, and $\mu_{r1} = \mu_{r2} = \mu_{r3} = 1$ (b) Decomposition of the geometry of Fig. 3a into triangular elements.

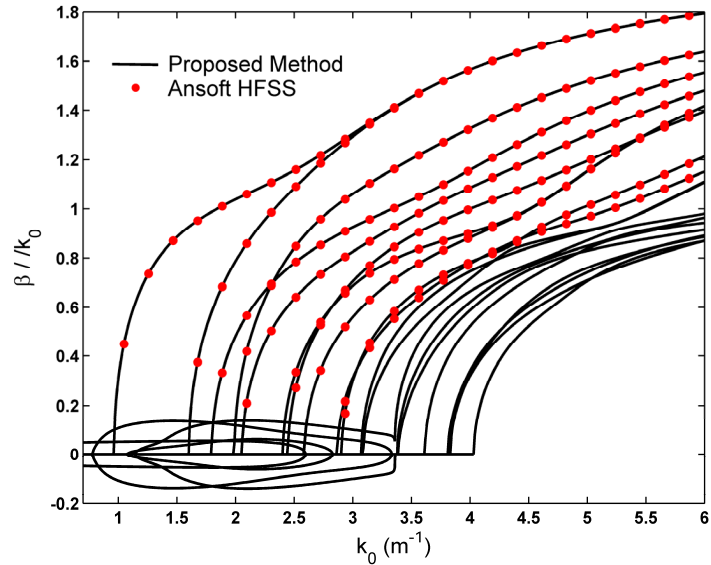

Fig. 4. Normalized propagation constants of the lower order modes of the waveguide of Fig. 3a versus frequency.

Table I Computational time, memory requirement, and total number of unknowns in the calculation of the first twenty modes of the waveguide of Fig. 3a at $k_0 = 3 \text{ m}^{-1}$.

| | Computational time (s) | | | | | | |
|---|---|---|---|---|---|---|---|
| | Evaluation of the matrices of all the elements | Assembly to construct the global matrix | Solving final matrix eigenvalue equation | Total computational time | Memory requirement (MB) | Total number of unknowns | Average relative error of the first ten modes |
| $M_u = 2, M_\varphi = 2$ | 0.42 | 0.05 | 0.22 | 0.75 | 0.25 | 437 | $9.3 \times 10^{-2}$ |
| $M_u = 4, M_\varphi = 4$ | 0.50 | 0.07 | 0.29 | 0.86 | 2.7 | 1745 | $1.3 \times 10^{-3}$ |
| $M_u = 6, M_\varphi = 6$ | 0.60 | 0.16 | 0.64 | 1.4 | 12 | 3917 | $8.7 \times 10^{-5}$ |
| $M_u = 8, M_\varphi = 8$ | 0.72 | 0.41 | 1.38 | 2.51 | 36 | 6953 | $8.1 \times 10^{-6}$ |
| $M_u = 10, M_\varphi = 10$ | 0.87 | 0.89 | 3.3 | 5.06 | 84 | 10853 | $4.9 \times 10^{-7}$ |
| $M_u = 12, M_\varphi = 12$ | 1.1 | 1.66 | 7.7 | 10.46 | 170 | 15617 | $3.9 \times 10^{-8}$ |
| $M_u = 14, M_\varphi = 14$ | 1.4 | 3 | 15.3 | 19.7 | 310 | 21245 | $3.1 \times 10^{-9}$ |
| $M_u = 16, M_\varphi = 16$ | 1.74 | 5.1 | 27.1 | 34 | 510 | 27737 | $2.6 \times 10^{-10}$ |

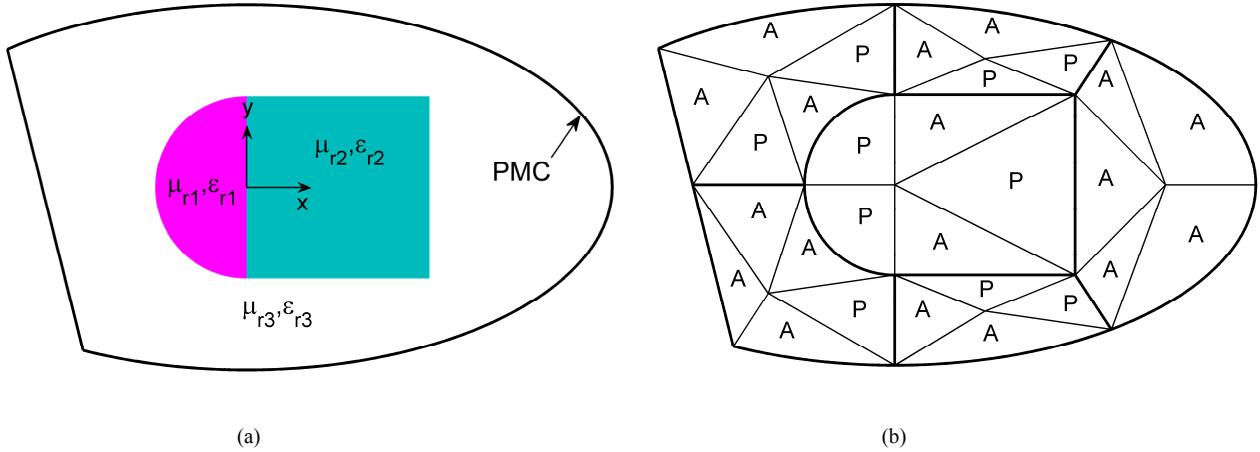

(a)                                                                (b)

Fig. 5. (a) Geometry of a singular waveguide, which is considered in section IV. The semicircle is defined by $x^2 + y^2 = 0.5^2$, where $x \le 0$. The outer elliptic boundary is defined by $(x/2)^2 + y^2 = 1$, where $-3\pi/4 \le \varphi \le 5\pi/6$. The side of the square is 1 m long. $\varepsilon_{r1} = 4$, $\varepsilon_{r2} = 2$, $\varepsilon_{r3} = 1$, and $\mu_{r1} = \mu_{r2} = \mu_{r3} = 1$ (b) Decomposition of the geometry of Fig. 5a into triangular elements.

## IV. NUMERICAL RESULTS

In order to investigate the convergence and computational efficiency of the proposed method, we first consider two distinct waveguide problems; the problem of a waveguide with a smooth geometry and regular field distribution and the problem of a waveguide including sharp corners and field singularities.

The geometry of the regular waveguide is shown in Fig. 3a and its decomposition into triangular elements of the type described in section II is shown in Fig. 3b. Also in Fig. 3b, we specify the boundaries of the curved polygons with the thicker lines and the principal and associate elements with the capital letters $P$ and $A$ inside each element, respectively. Total number of triangular elements in Fig. 3b is 36. To validate the method of section II, we calculate the normalized propagation constants of the first twenty modes of the waveguide of Fig. 3a versus frequency and represent the results in Fig. 4 along with those of Ansoft HFSS simulator. Excellent agreement which is observed between the both results in Fig. 4 confirms that the method is valid and is free of any spurious solution. In order to study the computational efficiency of the proposed method, we represent in Table I the computational time, memory requirement, and total number of unknowns in the calculation of the first twenty modes of the waveguide of Fig. 3a at $k_0 = 3 \text{ m}^{-1}$ and the average relative error of the first ten modes of the same waveguide for several values of $M_u$ and $M_\varphi$ inside the elements. For convenience, we assume the same $M_u$ and $M_\varphi$ for all the elements. It is clear from Table I that the evaluation of the matrices of all the elements through the 1-D integrations occupies a small portion of the total computational time. It should be noticed that the reduction of the 2-D integrals of the matrix elements to 1-D ones occurs inside the homogeneous elements or the elements whose electromagnetic properties are separable (i.e., can be represented as a product of two 1-D functions of $u$ and $\varphi$ where $(u, \varphi)$ are the local coordinates of the element).

In the next example, we investigate the accuracy and computational efficiency of the method in the analysis of a singular waveguide. The geometry of the waveguide is shown in Fig. 5a and its decomposition into 28 curved triangular elements is given in Fig. 5b. For this singular waveguide, the normalized propagation constants of the lower order modes versus frequency are calculated using the proposed method and depicted in Fig. 6 along with the results obtained from Ansoft HFSS simulator for comparison. Again, a good agreement is observed between the both results in Fig. 6 showing that the method is valid for singular waveguides. The average relative errors and computational times for the waveguide of Fig. 5a are calculated and listed in Table II. Table II again reveals the fact that the method is very fast in the evaluation of the elemental matrices. However, the error of the method for the singular waveguide is decreased efficiently up to a moderate value around $10^{-5}$ and further increasing the values of $M_u$ and $M_\varphi$ at this point, makes a little effect on the accuracy of the method.

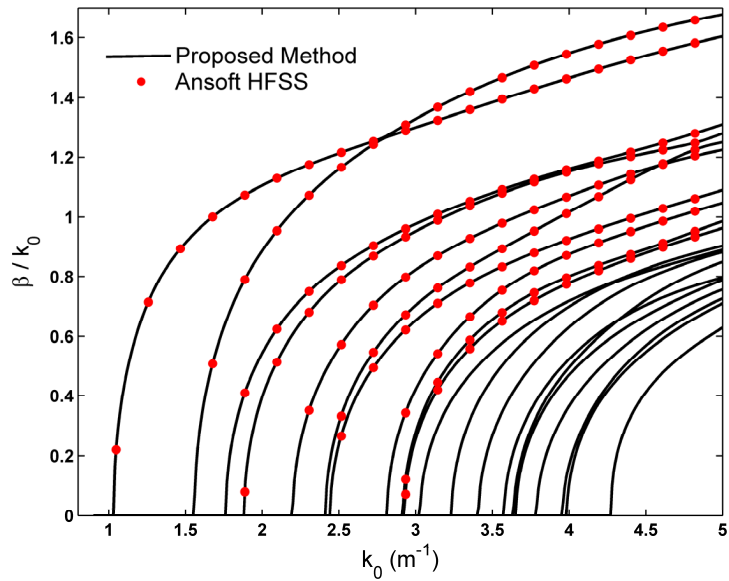

Fig. 6. Normalized propagation constants of the lower order modes of the waveguide of Fig. 5a versus frequency (complex modes are not shown for clearness).

Table II Computational time, memory requirement, and total number of unknowns in the calculation of the first twenty modes of the waveguide of Fig. 5a at $k_0 = 3 \text{ m}^{-1}$.

| | Computational time (s) | | | | Memory requirement (MB) | Total number of unknowns | Average relative error of the first ten modes |
|---|---|---|---|---|---|---|---|
| | Evaluation of the matrices of all the elements | Assembly to construct the global matrix | Solving final matrix eigenvalue equation | Total computational time | | | |
| $M_u = 2, M_\varphi = 3$ | 0.38 | 0.05 | 0.25 | 0.68 | 0.26 | 523 | $1.3 \times 10^{-1}$ |
| $M_u = 4, M_\varphi = 6$ | 0.45 | 0.08 | 0.4 | 0.93 | 4.3 | 2059 | $1.4 \times 10^{-3}$ |
| $M_u = 6, M_\varphi = 9$ | 0.57 | 0.22 | 0.87 | 1.66 | 20 | 4603 | $1.6 \times 10^{-5}$ |
| $M_u = 8, M_\varphi = 12$ | 0.68 | 0.59 | 2.25 | 3.52 | 60 | 8155 | $6.9 \times 10^{-6}$ |
| $M_u = 10, M_\varphi = 15$ | 0.89 | 1.41 | 5.5 | 7.79 | 140 | 12715 | $3.3 \times 10^{-6}$ |
| $M_u = 12, M_\varphi = 18$ | 1.18 | 2.71 | 9.5 | 13.36 | 285 | 18283 | $1.3 \times 10^{-6}$ |
| $M_u = 14, M_\varphi = 21$ | 2 | 5 | 19.2 | 26.2 | 520 | 24859 | $6.2 \times 10^{-7}$ |

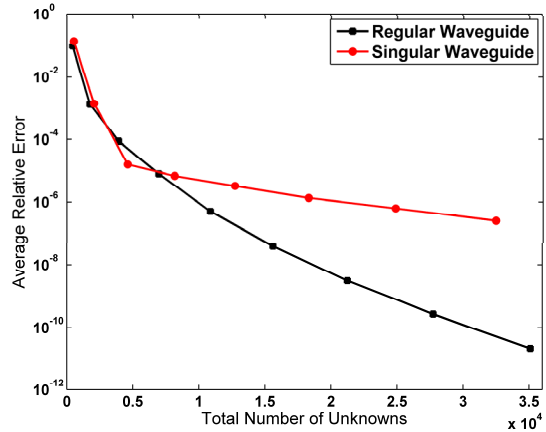

Fig. 7. Average relative errors of the first ten modes of the waveguides of Figs. 3a and 5a at $k_0 = 3$ m$^{-1}$ with respect to the total number of unknowns.

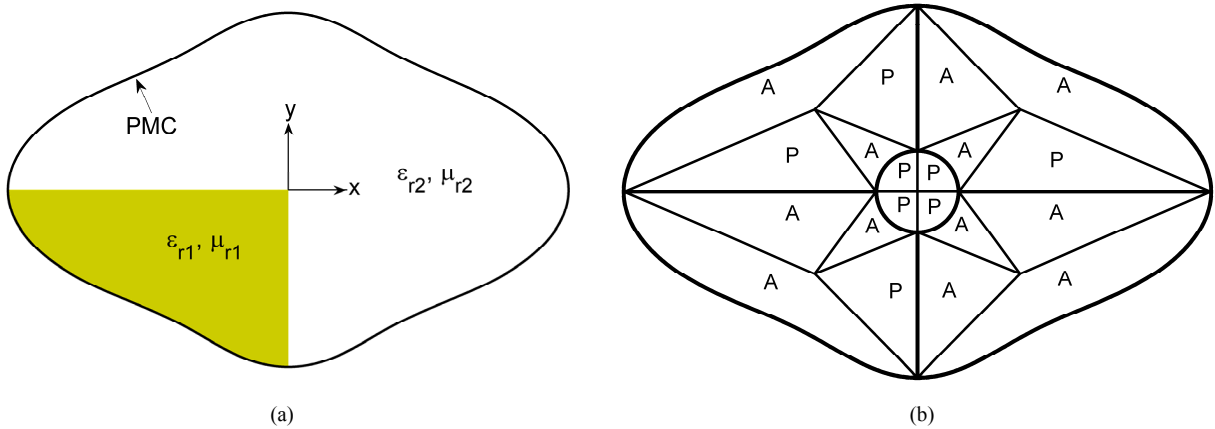

(a)                                                                 (b)

Fig. 8. (a) Geometry of a singular waveguide with a singular field point at its center. The boundary of the waveguide is defined by $\rho = 1/\bigl(1 - 0.2\cos(2\varphi) - 0.1\cos(4\varphi)\bigr)$, $\varepsilon_{r1} = 16$, $\varepsilon_{r2} = 1$, and $\mu_{r1} = \mu_{r2} = 1$ (b) Decomposition of the geometry of Fig. 8a into triangular elements.

Finally, we consider the convergence of the method with respect to the total number of unknowns. The convergence curves for the waveguides of Figs. 3a and 5a are depicted in Fig. 7. As can be seen from Fig. 7, the convergence is exponential for the regular waveguide while for the singular waveguide the convergence is exponential up to moderate accuracies and then is broken down and is fallen into an algebraic convergence. Therefore, when a higher accuracy is required, a special technique for the treatment of the singularity should be performed (e.g., h-refinements toward the singular points [21]). In order to improve the accuracy and convergence behavior of the method for singular waveguide problems, we apply a specific h-refinement that is compatible with our proposed finite element method. To be specific, we consider the waveguide of Fig. 8a including only one singular point, which is located at its center. The decomposition of the waveguide into curved polygons and triangles is shown in Fig. 8b, where a circular element centered at the singular point is considered. The circular element is divided into a number of segments in the radial direction, denoted by $S_u$, and the previously defined Chebyshev polynomials are used at each segment, while the definition of the bases in the peripheral direction remains unchanged. Now, the h-refinement can be imposed by applying a non-uniform segmentation that is refined toward the singular point (i.e., the center of the circle) and making use of lower order polynomials at the finer segments. The average relative error of the first ten modes of the waveguide of Fig. 8a versus total number of unknowns is represented in Fig. 9 for different number of segments in the singular circle. Fig. 9 shows that the accuracy and convergence behavior of the method is improved and the exponential convergence is recovered for the singular waveguide as the number of segments is increased in the singular element.

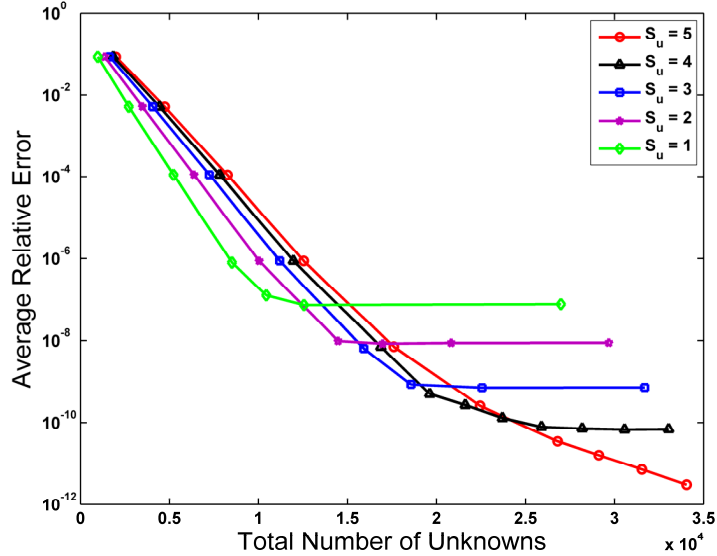

Fig. 9. Average relative error of the first ten modes of the waveguide of Fig. 8a at $k_0 = 3$ m$^{-1}$ with respect to the total number of unknowns for different number of segments in the singular circle.

## V. Conclusion

An efficient finite-element method has been proposed for the analysis of waveguides filled with pieces of homogeneous materials. The method divides the 2-D waveguide geometry into a small number of exact curved triangular elements and applies higher order hierarchical vector basis functions inside each element. The convergence and computational efficiency of the method have been studied through the analysis of a regular and a singular waveguide. It has been shown that the method is very fast and accurate for regular waveguides while it is moderately accurate for singular waveguides. For a higher accuracy in a singular waveguide a special technique for the treatment of the singularity should be added to the proposed algorithm.

## Appendix

The elements of the U sub-matrices in (2)-(10) are given by

$$\mathrm{U}^r_{m_1 m_2} = \int_0^1 F^r_{m_1 m_2} du \qquad (12)$$

where

$$F^1_{m_1 m_2} = \frac{U_{m_1}(2u-1)U_{m_2}(2u-1)}{u}, \qquad F^2_{m_1 m_2} = u\, U_{m_1}(2u-1)U_{m_2}(2u-1)$$

$$F^3_{m_1 m_2} = \frac{U_{m_1}(2u-1)T'_{m_2}(2u-1)}{u}, \qquad F^4_{m_1 m_2} = U_{m_1}(2u-1)T_{m_2}(2u-1)$$

$$F^5_{m_1 m_2} = u\, U_{m_1}(2u-1)T'_{m_2}(2u-1), \qquad F^6_{m_1 m_2} = \frac{T'_{m_1}(2u-1)T'_{m_2}(2u-1)}{u}$$

$$F^7_{m_1 m_2} = \frac{T_{m_1}(2u-1)T_{m_2}(2u-1)}{u}, \qquad F^8_{m_1 m_2} = T_{m_1}(2u-1)T'_{m_2}(2u-1)$$

$$F^9_{m_1 m_2} = u\, T'_{m_1}(2u-1)T'_{m_2}(2u-1), \qquad\qquad F^{10}_{m_1 m_2} = u\, T_{m_1}(2u-1)T_{m_2}(2u-1)$$

where $'$ represents the derivative with respect to $u$. The elements of the $\Phi$ sub-matrices in (2)-(10) are given by

$$\Phi^r_{n_1 n_2} = \int_{\varphi_1}^{\varphi_2} G^r_{n_1 n_2}\, d\varphi \qquad (13)$$

where

$$G^1_{n_1 n_2} = \frac{1}{\rho_1^2} T'_{n_1}(\psi)T'_{n_2}(\psi), \qquad G^2_{n_1 n_2} = \left(1 + \left(\frac{\rho'_1}{\rho_1}\right)^2\right)T_{n_1}(\psi)T_{n_2}(\psi), \qquad G^3_{n_1 n_2} = \frac{1}{\rho_1^2}T'_{n_1}(\psi)U_{n_2}(\psi)$$

$$G^4_{n_1 n_2} = \left(-\frac{\rho'_1}{\rho_1}\right)T_{n_1}(\psi)U_{n_2}(\psi), \qquad G^5_{n_1 n_2} = \left(-\frac{\rho'_1}{\rho_1}\right)T_{n_1}(\psi)T'_{n_2}(\psi), \qquad G^6_{n_1 n_2} = \frac{1}{\rho_1^2}U_{n_1}(\psi)U_{n_2}(\psi)$$

$$G^7_{n_1 n_2} = U_{n_1}(\psi)U_{n_2}(\psi), \qquad\qquad G^8_{n_1 n_2} = U_{n_1}(\psi)T'_{n_2}(\psi)$$

$$G^9_{n_1 n_2} = T'_{n_1}(\psi)T'_{n_2}(\psi), \qquad\qquad G^{10}_{n_1 n_2} = \rho_1^2\, T_{n_1}(\psi)T_{n_2}(\psi)$$

where $'$ represents the derivative with respect to $\varphi$ and $\psi = 1 - 2(\varphi - \varphi_2)/(\varphi_1 - \varphi_2)$.

For the associate elements, the integrands of the $\Phi$ sub-matrices in the above relations should be replaced by the following relations

$$G^1_{n_1 n_2} = \left(\frac{1}{\rho_2^2}\frac{d\varphi}{d\varphi'}\right)T'_{n_1}(\psi)T'_{n_2}(\psi), \qquad G^2_{n_1 n_2} = \left(\frac{\rho_1^2 + \rho'^2_1}{\rho_2^2}\frac{d\varphi}{d\varphi'}\right)T_{n_1}(\psi)T_{n_2}(\psi), \qquad G^3_{n_1 n_2} = \left(\frac{1}{\rho_2^2}\frac{d\varphi}{d\varphi'}\right)T'_{n_1}(\psi)U_{n_2}(\psi)$$

$$G^4_{n_1 n_2} = \left(-\frac{\rho'_2}{\rho_2}\frac{d\varphi}{d\varphi'}\right)T_{n_1}(\psi)U_{n_2}(\psi), \qquad G^5_{n_1 n_2} = \left(-\frac{\rho'_2}{\rho_2}\frac{d\varphi}{d\varphi'}\right)T_{n_1}(\psi)T'_{n_2}(\psi), \qquad G^6_{n_1 n_2} = \left(\frac{1}{\rho_2^2}\frac{d\varphi}{d\varphi'}\right)U_{n_1}(\psi)U_{n_2}(\psi)$$

$$G^7_{n_1 n_2} = \frac{d\varphi}{d\varphi'}U_{n_1}(\psi)U_{n_2}(\psi), \qquad\qquad G^8_{n_1 n_2} = \frac{d\varphi}{d\varphi'}U_{n_1}(\psi)T'_{n_2}(\psi)$$

$$G^9_{n_1 n_2} = \frac{d\varphi}{d\varphi'}T'_{n_1}(\psi)T'_{n_2}(\psi), \qquad\qquad G^{10}_{n_1 n_2} = \left(\rho_2^2\frac{d\varphi'}{d\varphi}\right)T_{n_1}(\psi)T_{n_2}(\psi)$$

where $'$ represents the derivative with respect to $\varphi$ and $\rho_2$, $\rho'_2$, and $d\varphi/d\varphi'$ can be calculated by the following formulas

$$\rho_2 = \sqrt{(x_c - x_a)^2 + (y_c - y_a)^2}, \quad \rho'_2 = \frac{x_c - x_a}{\rho_2}\frac{d(\rho_1 \cos(\varphi))}{d\varphi} + \frac{y_c - y_a}{\rho_2}\frac{d(\rho_1 \sin(\varphi))}{d\varphi}, \quad \frac{d\varphi}{d\varphi'} = \frac{\rho_2}{\sqrt{\rho_1^2 + \rho'^2_1 - \rho'^2_2}}$$

where $x_c = x_p + \rho_1 \cos(\varphi)$ and $y_c = y_p + \rho_1 \sin(\varphi)$.